\documentclass{aastex63}

\def\arcsec{^{\prime\prime}}
\def\arcmin{^{\prime}}
\def\degr{^\circ}
\def\hour{^{\rm h}}
\def\minute{^{\rm m}}
\def\second{^{\rm s}}

\received{--}
\revised{--}
\accepted{--}
\submitjournal{AJ}

\shorttitle{VZ Sex in $X$-rays}
\shortauthors{Nucita et al.}
\graphicspath{{./}{figures/}}
\begin{document}

\title{VZ Sex: X-ray confirmation of its intermediate polar nature}
\correspondingauthor{Achille A. Nucita}
\email{nucita@le.infn.it}
\author{A.A. Nucita}
\affiliation{Department of Mathematics and Physics {\it ``E. De Giorgi''} , University of Salento, Via per Arnesano, CP-I93, I-73100, Lecce, Italy}
\affiliation{INFN, Sezione di Lecce, Via per Arnesano, CP-193, I-73100, Lecce, Italy}
\affiliation{INAF, Sezione di Lecce, Via per Arnesano, CP-193, I-73100, Lecce, Italy}

\author{F. De Paolis}
\affiliation{Department of Mathematics and Physics {\it ``E. De Giorgi''} , University of Salento, Via per Arnesano, CP-I93, I-73100, Lecce, Italy}
\affiliation{INFN, Sezione di Lecce, Via per Arnesano, CP-193, I-73100, Lecce, Italy}
\affiliation{INAF, Sezione di Lecce, Via per Arnesano, CP-193, I-73100, Lecce, Italy}

\author{D. Licchelli}
\affiliation{R.P. Feynman Observatory, I-73034, Gagliano del Capo, Lecce, Italy }
\affiliation{CBA, Center for Backyard Astrophysics - I-73034, Gagliano del Capo, Lecce, Italy}

\author{F. Strafella}
\affiliation{Department of Mathematics and Physics {\it ``E. De Giorgi''} , University of Salento, Via per Arnesano, CP-I93, I-73100, Lecce, Italy}
\affiliation{INFN, Sezione di Lecce, Via per Arnesano, CP-193, I-73100, Lecce, Italy}
\affiliation{INAF, Sezione di Lecce, Via per Arnesano, CP-193, I-73100, Lecce, Italy}

\author{G. Ingrosso}
\affiliation{Department of Mathematics and Physics {\it ``E. De Giorgi''} , University of Salento, Via per Arnesano, CP-I93, I-73100, Lecce, Italy}
\affiliation{INFN, Sezione di Lecce, Via per Arnesano, CP-193, I-73100, Lecce, Italy}
\affiliation{INAF, Sezione di Lecce, Via per Arnesano, CP-193, I-73100, Lecce, Italy}

\author{M. Maiorano}
\affiliation{Department of Mathematics and Physics {\it ``E. De Giorgi''} , University of Salento, Via per Arnesano, CP-I93, I-73100, Lecce, Italy}
\affiliation{INFN, Sezione di Lecce, Via per Arnesano, CP-193, I-73100, Lecce, Italy}

\begin{abstract}
Intermediate polars are members of the cataclysmic variable binary stars. They are characterized by a moderately magnetized white dwarf accreting matter from a cool main-sequence companion star. In many cases, this accretion gives rise to a detectable $X$-ray emission.
   VZ Sex is an interesting $X$-ray source whose nature needs a robust confirmation. Here, we used archive $XMM$-Newton observation to assign the source to the intermediate polar class.
   We applied the Lomb-Scargle periodogram method to detect any relevant periodic feature in the $0.1$--$10$ keV light curve and performed a spectral fitting of the $X$-ray spectrum in order to get information on the on-going accretion mechanism.
   By inspecting the periodogram,  we detected a clear periodic feature at $\simeq 20.3$ minutes that we interpret as the spin period of the white dwarf. We additionally found the typical side bands expected as the consequence of the beat between the spin and the orbital period of $\simeq 3.581$ hours. The source is characterized by a unabsorbed flux of $\simeq 2.98\times 10^{-12}$ erg cm$^{-2}$ s$^{-1}$ corresponding to an intrinsic luminosity {of $\simeq 7 \times 10^{31}$ erg s$^{-1}$ } {for a  distance  of $\simeq 433$ pc}. The existence of such features allow us to classify VZ Sex as a clear member of the intermediate polar class.  Furthermore,  with  the  estimated  WD  spin,  the  ratio $P_{spin}/P_{orb}$ is $\simeq 0.09$, i.e.   consistent with that expected for a typical IP system above the period gap.  In addition, the estimated intrinsic luminosity opens the possibility that a bridge linking the normally bright IPs to the faint population of sources does exist.
   
\end{abstract}


%

\section{Introduction}

Intermediate polars (hereinafter IPs) are a class of cataclysmic variables (CVs) characterized by a spinning, moderately magnetized (with magnetic field intensity in the range $0.1$ -- $10$ MG) white dwarf (WD) accreting material from a donor companion. In general, CVs characterized by a magnetic field strength $\lesssim 0.1$ MG are known as {\it dwarf novae} (see, e.g. \citealt{vanteeseling1996,nucita2009,hoard2010,nucita20092,nucita2011,balman2011,nucita2014,mukai2017}) {while systems with magnetic} field exceeding $10$ MG are instead called {\it polars} (see, e.g., \citealt{ramsay2004, szkody2004}). The reader is referred to  \cite{Kuulkers2006} for a review.

In these binary systems the material initially circulates in an accretion disk that is then truncated by the magnetic field of the WD and a {\it curtain} forms with the stream of matter falling on the WD poles. In the accretion column, the matter suffers a shock wave and, consequently, releases $X$-ray photons whose spectrum strongly depends on the details of the accretion process. 

In IPs, the high energy signal is modulated on the WD spin $P_{spin}$ and on the orbital period $P_{orb}$ (\citealt{parker2005}). However, as observed by 
\cite{warnerbook} (but see also \citealt{king1992}), since the $X$-ray production source shows a variable sight to the observer and due to the presence of reprocessing sites, multiple orbital side-bands are expected and, often, observed. For example, it is expected to observe periodic features at the frequencies $P_{spin}^{-1}\pm P_{orb}^{-1}$ and $P_{spin}^{-1}\pm 2P_{orb}^{-1}$ due to amplitude modulation at $P_{orb}$ and $2P_{orb}$. 

As noted by \cite{nucita2019, nucita2020}, but see also the analysis by \cite{mukai2020}, the existence of multiple peaks in the Fourier transform of the high energy light curve offers a powerful tool to assign, routinely, the IP identification to many CV systems. {Therefore, dedicated observations in the  $X$-ray band} with large sensitivity in the $0.1$-$15$ keV range (as that offered by the $XMM$-Newton satellite onboard instruments) are crucial for a correct classification of such objects.

Although IPs are expected to be common in the Galaxy (see, e.g.,  \citealt{worrall1982}) and to contribute substantially to the diffuse $X$-ray ridge background \citep{revnivtsev2009,warwick2014}, only for a restricted number of sources the IP nature was claimed (see e.g. the updated intermediate polar catalogue -{\it IPhome}- available at \url{https://asd.gsfc.nasa.gov/Koji.Mukai/iphome/iphome.html}). Indeed, to that aim, dedicated timing and spectral analyses are necessary for each candidate.

Here,  we  continue  our  project  aimed  to  the  classification of  IP  candidates  based  on  a  detailed $X$-ray  data  analysis. A class study of the identified IPs will be presented later elsewhere. In particular, in this paper we concentrate  on  the  CV source  VZ  Sex  (also  known  as  1RXSJ094432.1+035738) which is recognized to be an IP candidate with an orbital period\footnote{In this work, based on the results of the timing analysis reported in Section 3, we assume for the VZ Sex orbital period the value of $\simeq 3.581$ hours \citep{mennickent2002} (but see also the the updated IPhome catalogue). We note that this value is slightly larger than the period estimates of $3.570$--$3.576$ hours \citep{gansicke2009} and $3.562$--$3.576$ hours \citep{thorstensen2010}.} of $\simeq 3.581$ hours (see also \citealt{mennickent2002}).

At present, no clear identification of the WD spin is known in the literature, although a tentative period of $\simeq 40.83$ minutes is reported in the IPhome catalogue. Furthermore,  \citet{mennickent2002} tentatively classify the source as a U Geminorum like dwarf novae although the IP nature cannot be {excluded} by the same authors. We show that the high energy light curve of this source is characterized by a modulation on $\simeq 20.3$ minutes that we interpret as the spin period of the WD. This is supported by the fact that the typical IP side-bands, forming when the spin period beats with the orbital period $P_{orb}\simeq 3.581$ hours, are found in the Lomb-Scargle periodogram. The source is also characterized by a $0.1-10$ keV spectrum which is well described by a multi-temperature plasma model ({based on} the cooling of the gas when falling onto the WD surface) absorbed by the galactic neutral hydrogen column density. The estimated unabsorbed flux is $\simeq 2.98\times 10^{-12}$ erg cm$^{-2}$ s$^{-1}$ corresponding to an intrinsic luminosity of $\simeq 8.4\times 10^{31}$ erg s$^{-1}$  for a {distance\footnote{
 Note here that the distance estimated by \citet{mennickent2002} is rather uncertain and one expects that Gaia data would lead to a much better distance determination. Unfortunately, Gaia DR2 data \citep{gaiadr2} does not report any parallax value for VZ Sex so we will assume $433\pm 100$ pc as the distance to the target.} to the source of $433\pm 100$ pc \citep{mennickent2002}.} All that allows us to conclude that VZ Sex is certainly an IP cataclysmic variable star.

The outline of the paper is as follows: in Sections 2 and 3 we summarize the adopted X-ray data reduction procedure and the performed timing analysis. In Section 4 we {present} the X-ray spectral analysis {along with} our results and, {in Section 5, we finally discuss the results.}

\section{X-ray data reduction}
The cataclysmic variable VZ Sex (also known as 1RXS J094432.1+035738) is located at J2000 coordinates ${\rm RA=09\hour 44\minute 31.72\second}$  and ${\rm DEC = +03\degr 58\arcmin 05.4\arcsec}$. 
{The target was first pinpointed by means of the {\it Röntgen Satellite} and reported in the ROSAT All-Sky Bright Source Catalogue \citep{rosatcatalogue} as a source having a $0.1-2$ keV count rate of $0.087\pm0.022$ count s$^{-1}$  (see Section 5 for further details).}

{In order to study any IP nature of the source, VZ Sex was then pointed  by the $XMM$-Newton satellite in 2004 (Observation ID 0201290301, P.I. De Martino\footnote{{For completeness, we note here that VZ Sex and  two additional IP candidates (UU Col and RXJ1039-0507) were the targets of an {\it XMM}-Newton proposal associated to several observations (with Observation IDs 0201290101, 0201290201, 0201290301, and 0201290401). The observations were conducted in order to exploit the improved $XMM$-Newton capabilities with respect to those of previous satellites. In the case of UU Col, \citet{demartino2006} clearly demonstrated that the source is an IP since the high energy ($0.2-15$ keV) light curve is characterized by a  $\simeq 863$ seconds periodicity (identified as the WD spin period). Weak variabilities at the beat $935$ seconds and at the $3.5$ hours orbital periods were also observed with the orbital modulation being more evident  in the soft energy band (below $0.5$ keV). As far as the IP RXJ1039-0507 is concerned,   \citet{woudt2003} reported an orbital period of $1.574$ hours and found periodic features at $1932.5$ seconds and $721.9$ seconds which were interpreted as the sideband and first harmonic of the WD spin period of $1444$ seconds, respectively. Here, we concentrate on VZ Sex whose IP nature is still unclear.}})} with the observation starting (ending) on $2004/05/18$ at $14$:$27$:$42$ UT ($2004/05/19$ at $21$:$24$:$20$ UT), i.e. for a nominal observing time of $\simeq 37.6$ ks. The target was observed by all the instruments on board the satellite and, in particular, with the MOS 1, MOS 2 and pn cameras operated in full frame mode and with the thin filter on.  

We processed the observation data file (ODFs) by using the $XMM$-Science Analysis System {(SAS version 17.0.0, \citealt{gabriel2004})} with the most updated calibration constituent data files (CCFs). After we run the SAS tasks {\it emchain} and {\it epchain}, we were left with the calibrated event files. We further corrected (via the {\it barycen} SAS tool) the arrival times of the $X$-ray photons to account for the Solar System barycenter and accounted for the out-of-time events by following the recipes described in \cite{xrp}. In order to reduce the soft proton background possibly affecting the data, we built light curves of all the cameras considering only photons with energy above $10$ keV. By requiring that the net count rate is below $0.4$ count s$^{-1}$, we flagged the good time intervals (GTIs) and cut the event list files accordingly. In the case of VZ Sex the observation was affected by strong flares for $\simeq 60\%$ of the whole duration in the second half of the observation. Hence, in the following analysis, we considered only the events falling in the first 20 ks. The event list files corrected for the flares were then used to generate the science products as the light curves in the soft ($0.1$--$2$ keV), hard ($2$--$10$ keV) and full ($0.1$--$10$ keV) and images in the $0.1-10$ keV band for each camera for inspecting purposes. In all the cases, the source plus background signal was extracted on a circular region centred on the nominal position of the target and with a radius of $\simeq 40\arcsec$ ensuring that  $\simeq 90\%$ of the total energy of the source is correctly collected. The background extraction regions were positioned close to VZ Sex, (when possible) on the same chip, and avoiding to encircle any other visible source. 

\begin{figure*}
  \centering
{\includegraphics[width=0.9\textwidth]{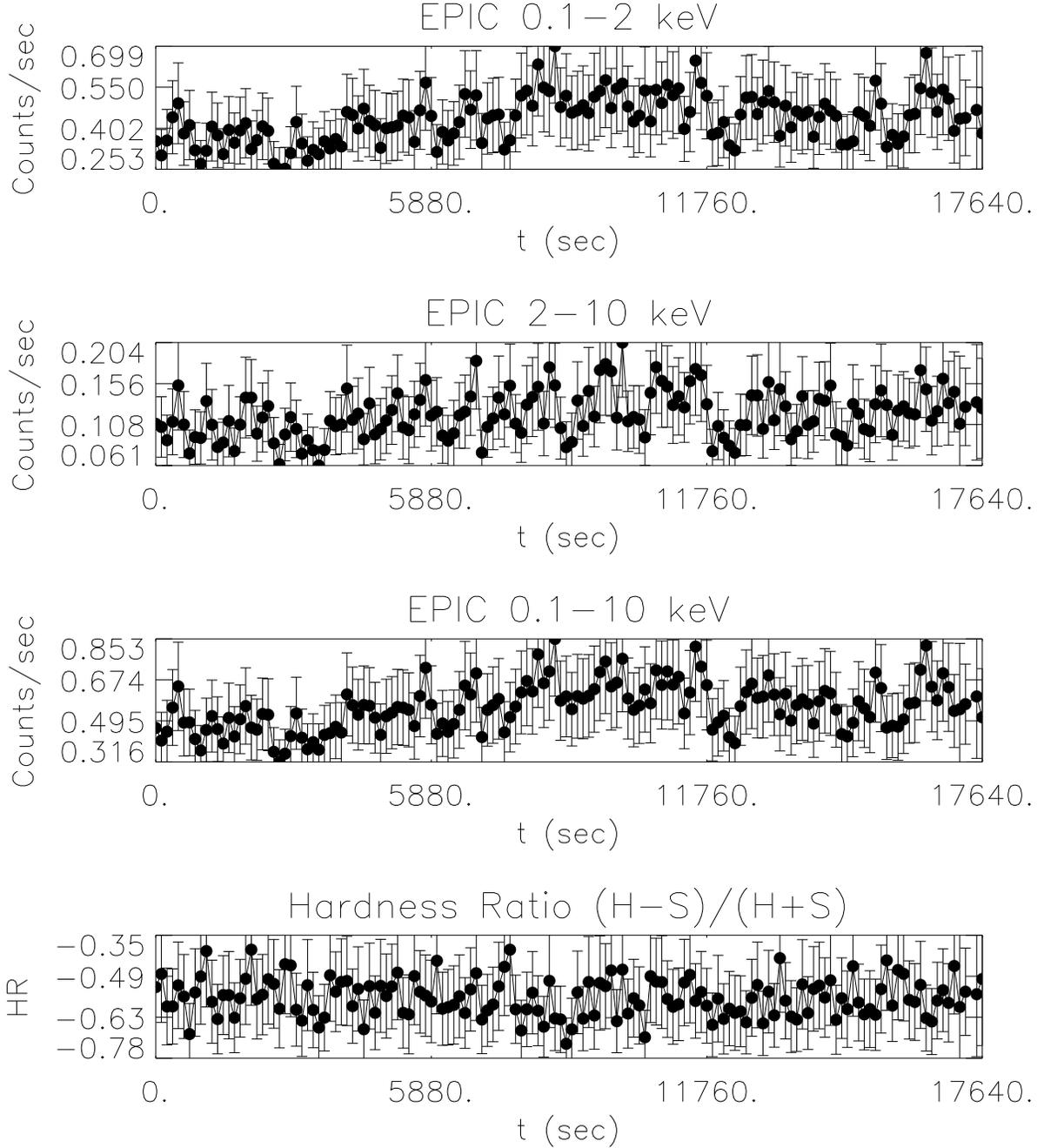}}
\caption{From top to bottom: the VZ Sex light curves in the $0.1-2$ keV, $2-10$ keV, $0.1-10$ keV (background subtracted and synchronized)  and the hardness ratio time series (see text for details). {The bin size was fixed to 120 seconds.} }
\label{fig_l}
\end{figure*}

For each camera and band (soft, hard, and full) we extracted the source time series {corresponding  to  the overlapping time intervals} and flagged the starting and stopping times of the MOS 1, MOS 2 and pn events. Then, we produced in each band synchronized\footnote{Synchronizing the source and background light curves means that each time series starts and ends exactly at the same instants of time and have the same  bin size. As a consequence, the background subtraction can be done on a bin-by-bin basis.} light curves with a bin size of $10$ seconds {(for the subsequent analysis) and $120$ seconds (for graphical purposes only)}. We followed the same procedure for the data contained within the background regions so that we have, for each camera and band, synchronized source (plus noise) and background light curves having the same bin size. The source (background corrected) light curves are then obtained by using the {\it epiclccorr} task that accounts for the different areas and exposures corrections. The final MOS 1, MOS 2 and pn source (background subtracted) were then averaged bin-by-bin and scaled in order to start from $0$. In Figures \ref{fig_l} we give the soft, hard, full and hardness ratio (see next Section for details) $X$-ray light curves of VZ Sex with a bin size of 120 seconds.
Finally, we extracted the spectra for the source and background and rebinned the data by requiring to have at least  25 counts per energy bin. Furthermore, we determined the associated response matrices and ancillary files that were used within {the XSPEC software (version 12.9.0, \citealt{arnaud})} during the spectral analysis of  the source and the estimate of the $0.1-10$ keV band flux.

\section{X-ray timing analysis}
{The barycentric corrected, synchronized and background corrected light curves extracted  are shown in Figure \ref{fig_l} with a bin size oif 120 seconds. The combined light curves (averaged over the instruments) have average count rate of $0.43\pm0.16$ count s$^{-1}$, $0.12\pm0.07$ count s$^{-1}$, and $0.55\pm0.18$ count s$^{-1}$ for the soft, hard, and full bands, respectively.} Here, the start of the observation corresponds to $MJD=53143.622$ and the whole light curve lasts for $17.76$ ks. It is then clear that the source is soft, i.e. it has an emission in the soft band larger than that in the hard part of the spectrum. This is also observed from the hardness ratio light curve $HR$ (bottom panel in Figure \ref{fig_l}) defined as $HR=(H-S)/(H+S)$, where $S$ and $H$ are the count rate in the soft and hard band, respectively. Inspection of this figure shows that $HR$ remains negative (and almost constant) for all the duration of the observation.

For our analysis we extracted the soft, hard and full band light curves with bin size of $10$ seconds and searched for periodic features by using the Lomb-Scargle technique (\citealt{scargle1982}). In particular, we restricted our analysis to the range between $2 \Delta t$ (where $\Delta t$ is the adopted bin size) and one half of the observational window so that we can test periodicities up to $\simeq 140$ minutes. We stress that the VZ Sex orbital period is $P_{orb}\simeq 3.581$ hours, i.e. well above our upper limit to the testable period so that any periodic feature corresponding to $P_{orb}$ cannot be picked up by our analysis. Note that with the quoted orbital period, VZ Sex is an IP candidate well above the period gap of $2$--$3$ hours so that a ratio between the WD spin and the orbital period in the range $0.01$--$0.1$ is expected.
\begin{figure*}
  \centering
  {\includegraphics[width=0.9\textwidth]{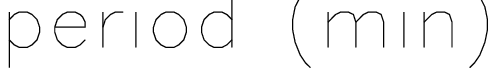}}
\caption{The Lomb-Scargle periodogram associated to the VZ Sex light curves in the $0.1-2$ keV (red line), $2-10$ keV (green line), and $0.1-10$ keV (black line) energy band. A $10$ second bin size has been used.  The inset shows a zoom of the Lomb-Scargle periodograms in the period range $2$--$25$ minutes. The meaning of the vertical lines is detailed in the text.}
\label{fig_lombscargle}
\end{figure*}
The result of the analysis is shown in Figure \ref{fig_lombscargle} where we give the Lomb-Scargle periodograms associated to the VZ Sex light curves in the $0.1-2$ keV, $2-10$ keV, and $0.1-10$ keV with red, green and black lines, respectively. A clear periodic feature at $20.3\pm 1.0$ minutes (labelled by the red dashed vertical line) is detected in the soft band with the signal that almost disappears in the hard part of the spectrum (as it usually happens in sources belonging to the IP sub-class) and obviously re-appears in the periodogram corresponding to the full light curve where
the count rate is still dominated by the $0.1-2$ keV photons.

We tentatively identify the detected periodicity with the WD spin and associate to it, as an error, the full  width at half  maximum  of the associated power peak. In this respect, note that we do not recover any strong feature at $\simeq 40$ minutes as reported in the {\it IPhome} site.
This conclusion is supported by the fact that, as also observed by \citealt{parker2005}, the $X$-ray signal coming from IP sources is often characterized by modulation on the WD spin ($P_{spin}$) and on the binary system orbital period ($P_{orb}$). In fact, as also explained by \citet{warnerbook}, the spinning of the WD polar caps (one of the production sites of $X$-ray photons) and the existence of other reprocessing loci induce the appearance of several side-bands whose periodogram peak intensities depend on the actual geometry of the system and on the line of sight to the observer (see, e.g. \citealt{king1992}). For example, one expects to find a feature at the synodic period   $P_{syn}^{-1}=P_{spin}^{-1} - P_{orb}^{-1}$  and also at the beat period $P_{beat}^{-1}=P_{spin}^{-1} + P_{orb}^{-1}$. Therefore, as stressed by \citet{nucita2019,nucita2020} (but see also \citealt{Joshi2016} and \citealt{mukai2020} for similar studies applied to strongly and moderately magnetized CVs) finding signatures of the orbital and spin period (along with the associated side-bands) is a robust method for the classification of a given source as an IP member. 

This is exactly the case of VZ Sex as indicated in the inset of Figure   \ref{fig_lombscargle} which shows a zoom of the Lomb-Scargle periodograms in the period range $2$--$25$ minutes. The blind search for modulations on the WD spin and the orbital period allowed us to detect the synodic feature at $P_{syn}\simeq 22.5$ minutes (orange dashed line) and the beat periodicity $P_{beat}\simeq 18.6$ minutes (blue dashed line) exactly at the expected locations in the assumption that the periodogram peak at $\simeq 20.3$ minutes corresponds to the WD spin. We also detected other multiple beat frequencies such as, in particular, the $P_{spin}^{-1}+3 P_{orb}^{-1}$ beat at $\simeq 15.8$ minutes,  green dashed line), the $2P_{spin}^{-1}+3 P_{orb}^{-1}$ beat at $\simeq 8.9$ minutes (red dotted line), the $3P_{spin}^{-1}+2 P_{orb}^{-1}$ beat at $\simeq 6.4$ minutes  (orange dotted line) and $3P_{spin}^{-1}-2 P_{orb}^{-1}$ beat at $\simeq 7.2$ minutes (blue dotted line). 
\begin{figure*}
  \centering
\includegraphics[width=0.9\textwidth]{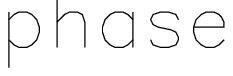}
\caption{The VZ Sex light curve is folded at the WD spin   
of $\simeq 20.3$ minutes with 20 bins per cycle. Here, the zero phase corresponds to the start of the observation at $MJD=53143.6192$.}
  \label{fig_folded}
\end{figure*}
With the estimated WD spin, the ratio $P_{spin}/P_{orb}$ results to be $\simeq 0.09$, i.e. consistent with what expected for a IP system above the period gap.

We then folded the $0.1-10$ keV light curve of VZ Sex at the detected WD spin $\simeq 20.3$ minutes with 20 bins per cycle (see Figure \ref{fig_folded}). {The resulting light curve shows a clear sinusoidal pattern with the zero phase corresponding to the start of the observation at $MJD=53143.6192$, possibly associated to a change on the projected emitting area.  We also note that the modulation is less evident at energies larger than 2 keV as a consequence of the lower count rate. Moreover, the folded light curve shows a possible dip at phase $\simeq 0.7$ which lasts for $\simeq 0.1$. We then test the possibility that the modulation is due to intervening absorption. However, as discussed in the next section, we find that the data at the phase peak and at the dip do not show any clear spectral change.} 

Therefore, based on the results presented in this section, VZ Sex can be safely classified as a member of the IP subclass (see, e.g. \citealt{nucita2019, nucita2020}, for similar recent analyses applied to DW Cnc, HP Cet and Swift J0820.6-2805).

\section{X-ray spectral analysis}
\begin{figure*} 
  \centering
  \includegraphics[width=0.7\textwidth, angle=-90]{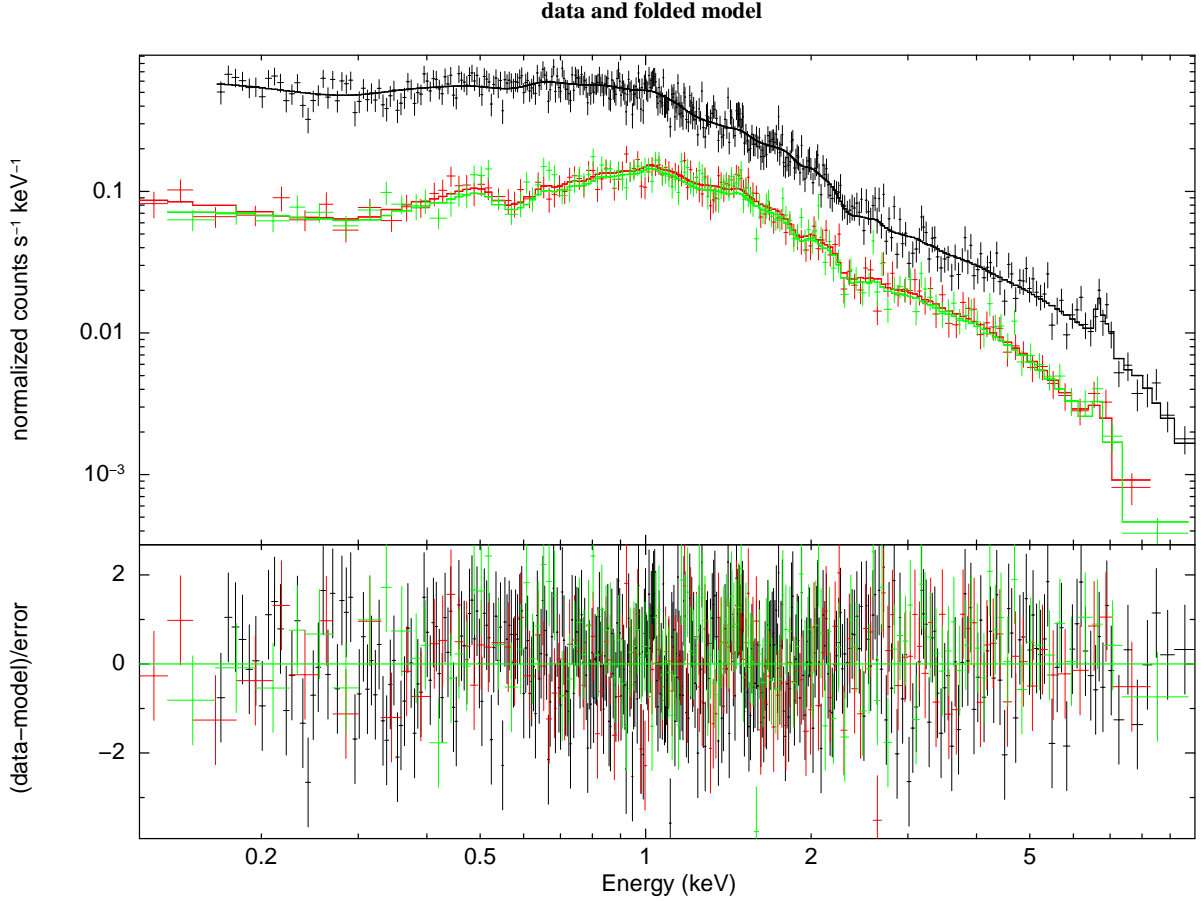}
  \caption{The $0.1-10$ keV spectra for the MOS 1 (red), MOS 2 (green) and pn (black) data of VZ Sex together with the best-fit model (see text for details) are presented.}
  \label{spectrum}
\end{figure*}

VZ Sex is a moderately bright source characterized by a $0.1$--$10$ keV band rate of $\simeq 0.55$ counts s$^{-1}$ which, for a total observation duration of $\simeq 20$ ks, corresponds to a sufficiently large number of collected photons ($\simeq 1.1\times 10^4$) distributed in a relatively high quality spectrum of the source. As previously discussed, the MOS 1, MOS 2, and pn spectra for the source were imported within the XSPEC package along with the spectra for the background and the response matrices  of the instruments (see, Figure \ref{spectrum} where the black, red and green points in the upper panel correspond to the pn, MOS 1, and MOS 2 data, respectively).
By using XSPEC, we started the analysis by fitting the data with a single mekal model \citep{mekal} absorbed by the neutral hydrogen {({\it phabs} component in XSPEC)} placed along the line of sight which, in this particular case, is $\simeq 3.84\times 10^{20}$ cm$^{-2}$ \citep{nhtool}. By keeping fixed the abundance to the solar value, the fit converged towards a solution characterized by $kT\simeq 5$ keV and a $n_H\simeq 1.75\times 10^{20}$ cm$^{-2}$ which is lower than that provided by the on-line {\it $n_H$ tool}\footnote{The tool is available at 
\url{https://heasarc.gsfc.nasa.gov/cgi-bin/Tools/w3nh/w3nh.pl}}, as expected from the fact that VZ Sex is located at a distance of $\simeq 433$ pc. However, the fit is formally unacceptable (having the reduced $\chi^2=1.4$ for 692 d.o.f.) and large residuals appear close to the iron line complex at $\simeq 6.5$ keV. Allowing the metal abundance $A$ to vary resulted in a model with $A\simeq 0.3$ and the other parameters practically unchanged. Although the residuals disappear, the fit is still not acceptable since it is characterized by  $\chi^2=1.2$ (reduced) for 691 d.o.f. {Therefore, there is the possibility that a multi-temperature plasma is acting. Hence, after adding a second mekal component, the fit converged towards the solution $n_H = (1.713\pm 0.002)\times 10^{20}$ cm$^{-2}$, $kT_1= 5.9^{+0.5}_{-0.3}$ keV, $kT_2= 0.70^{+0.02}_{-0.05}$ keV, and $A=0.36\pm 0.09$ with $\chi^2=1.06$ (reduced) for 689 d.o.f. All the errors are quoted at the 90\% confidence level.}

{In IP sources, it is expected that the accretion post shock regions are characterized by a temperature gradient due to the cooling of the infalling gas \citep{demartino2005}. We then used XSPEC {\it cemekl} model in which the plasma is  characterized by an emission measure gradient following a power law $dEM/dT=(T/T_{max})^{\alpha-1}/T_{max}$. Therefore, the free parameters of the model are the neutral hydrogen column density $n_H$, the maximum value of the temperature $T_{max}$ in the plasma, the power law index $\alpha$, the metal solar abundance $A$, and the model normalization $N$. For any given set of parameters, the spectrum is still dependent on the mekal model and obtained by interpolating on a pre-calculated mekal table (i.e. setting the {\it switch} parameter to 1. We verified that running the code by setting the {\it switch} parameter to 2 (i.e. by using the intrinsic {\it apec} model) does not change our results.}

With this model, the fit converged towards the solution $n_{H}=(1.990\pm 0.003)\times 10^{20}$ cm$^{-2}$, $kT_{max}= 13.7^{+2.7}_{-1.9}$ keV, $\alpha=0.9 \pm 0.2$, $A=0.3\pm 0.1$ {that is formally equivalent to the two-component mekal model described above with $\chi^2=1.02$ (reduced) for 690 d.o.f.} The addition of a partial covering factor did not improve the quality of the fit with respect to the cemekl model, so we do not consider  further the presence of a local absorption.

Considering the cemekl model as a good representation of the accretion process in the case of VZ Sex, we found that the  $0.1$--$10$ keV band absorbed flux is { $(2.67_{-0.5}^{+0.4})\times 10^{-12}$ erg cm$^{-2}$ s$^{-1}$}. The unabsorbed flux\footnote{{We note here that the {\it XMM}-Newton derived flux is consistent with that ($\simeq 3\times 10^{-12}$ erg cm$^{-2}$ s$^{-1}$) derived by converting the count rate of $\simeq 0.0876$ counts s$^{-1}$ of the ROSAT satellite observation (and reported on the ROSAT All-Sky Bright Source Catalogue, \citealt{rosatcatalogue}) into a flux in the $0.1-10$ keV band.}} is $(2.98_{-0.5}^{+0.4})\times 10^{-12}$ erg cm$^{-2}$ s$^{-1}$ so that, for the {VZ Sex distance of $433\pm 100$ pc}, the intrinsic source luminosity {results to be $(7 \pm 4)\times 10^{31}$ erg s$^{-1}$}. This result is in agreement with the distribution in luminosity of the secure IP sample observed in the Swift-BAT 70-month survey \citep{pretorius} although at the lower end of the distribution of the sub-sample with luminosity $\simeq 10^{33}$ erg s$^{-1}$ and at the upper end of the under-luminous IP (see \citealt{mukai2020} and \citealt{nucita2020} for a discussion on other interesting faint IPs). In this respect, VZ Sex seems to be on the bridge between the known bright IP distribution and the still poorly constrained faint IP sources.

\begin{figure*} 
  \centering
  \includegraphics[width=0.7\textwidth, angle=-90]{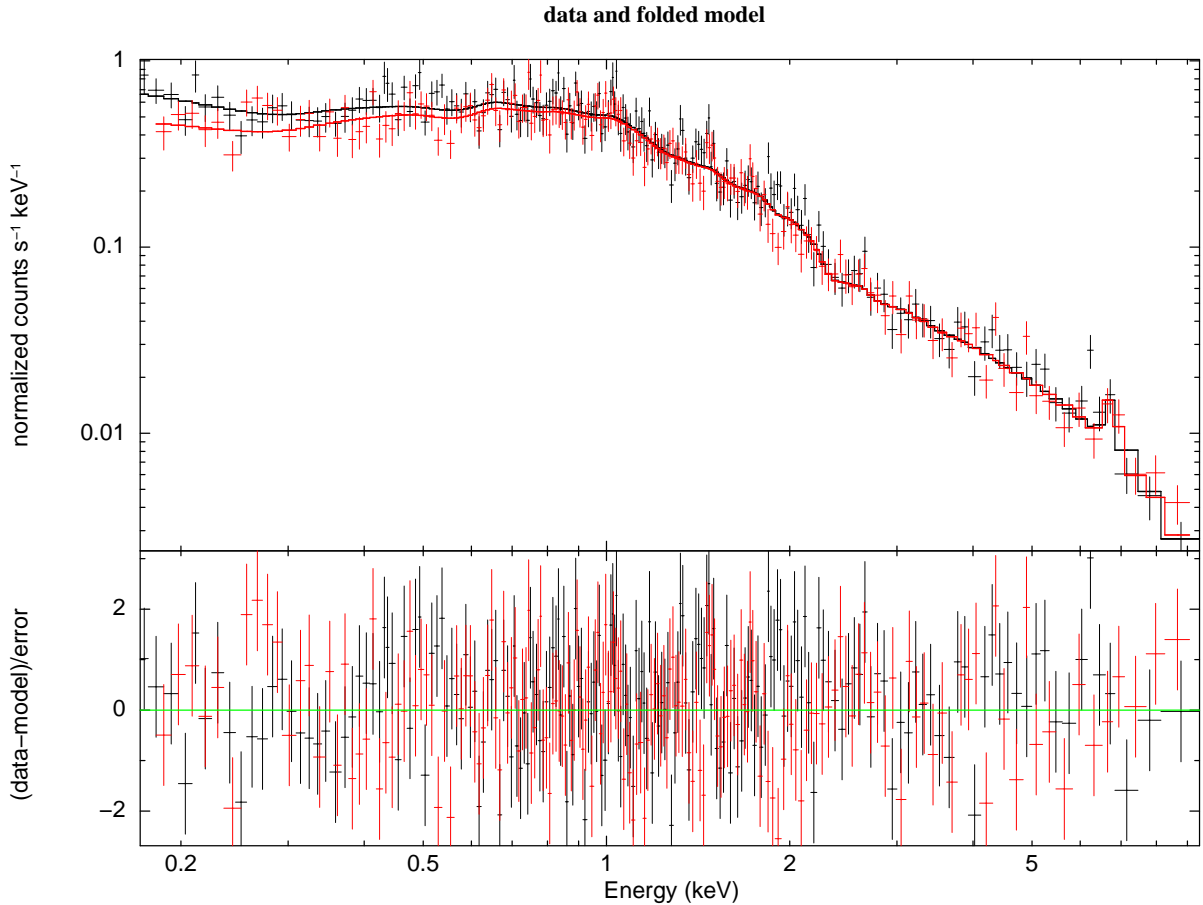}
  \caption{The pn $0.1-10$ keV spectra obtained by extracting the data around the maximum (black data) and minimum (red data) of the folded light curve.}
  \label{spectrum_folded}
\end{figure*}

{We further investigate the white dwarf spin pulse profile in order to find any possible dependence of the spectral properties on the WD spin phase. Hence, we first defined phases intervals corresponding to the region enclosing the minimum of the light curve (at phases $0\leq \phi \le 0.3$ and $0.9\leq \phi \le 0.1$), around the maximum (at phases $0.3 < \phi \leq 0.6$ and $0.8\leq \phi < 0.9$) and surrounding the possible dip (see Figure  \ref{fig_folded}) at phases $0.6 < \phi < 0.8$. By using the {\it phasecalc} task of the SAS suite, we calculated the phases associated to the collected $X$-ray events assuming that the zero phase corresponds to the starting time of the observation at $MJD= 53143.6192$. Hence, for each of the interesting phase intervals, we extracted the spectra (in Figure \ref{spectrum_folded} we give for clarity only the data corresponding to the pn camera). We fit the data by using the cemekl model and found that all the interesting fit parameters converged to values consistent with those obtained above. We then fixed all the parameters to the best values obtained for the full spectrum apart for the hydrogen column density which is free to vary among the data. In this case, we are able to verify whether the modulation observed in the folded light curve is due to an intrinsic absorption. The black and red data in Figure \ref{spectrum_folded} correspond to the data at maximum and minimum of the folded light curve, respectively. We note that the data at maximum and minimum are affected by a column density of $n_{H}=(1.633\pm 0.004)\times 10^{20}$ cm$^{-2}$ and $n_{H}=(2.295\pm 0.003)\times 10^{20}$ cm$^{-2}$, respectively, i.e. the photons at maximum are less absorbed than at minimum (and for energies below $\simeq 1$ keV), although the effect is not dramatic.  Conversely, the data associated to the possible dip does not show any change in the spectral change with respect to the data at maximum and, as a consequence, we do not show the corresponding data in Figure \ref{spectrum_folded}. We thus conclude that the modulation observed in the light curve is principally due to a change in the projected emitting surface. 
}

\section{Results and discussion}

Intermediate polars are cataclysmic variables in which a moderately magnetized (with magnetic field up to $\simeq 10$ MG) white dwarf accretes material from a companion star. The infalling material can be heated up to temperatures sufficient to produce $X$-ray photons up to energies of 20-50 keV. 

IPs are expected to be rather common in the Galaxy (\citealt{pretorius}) and it was shown that most of the confirmed IPs have a typical luminosity of $\simeq 10^{33}$ erg s$^{-1}$, but a sub-class of faint sources (as IP V597 Pup, V475 Sgr and  HP Cet, Swift J0820.6-2805) also exists. 
In any case, since IPs can be intrinsically faint, they are hard to be detected and, for each source, a dedicated $X$-ray observation is required in order to describe correctly the spectral features which strongly depend on the accretion mechanism. 
Furthermore, as noted by \citet{king1992}, \citet{warnerbook} and \citet{parker2005}, the $X$-ray signal from an IP source is characterized by modulation on the WD spin and on the binary system orbital period which pop up as extra peaks in the Fourier power spectrum. As stressed, e.g., by \citet{Joshi2016} (but see also \citet{mukai2020}, this offers a unique tool to confirm the nature of many IP candidates provided that long duration $X$-ray observation by large sensitivity instruments (as those on-board the $XMM$-Newton satellite) are available.

In this paper, we concentrated on the CV source VZ Sex (1RXS J094432.1+035738), an IP candidate with an orbital period (above the gap) of $\simeq 3.581$ hours. Here, by studying the soft ($0.1$--$2$ keV), hard ($2$--$10$ keV), and full ($0.1$--$10$ keV) band light curve we clearly identify in the associated periodograms a feature corresponding to the periodicity of $\simeq 20.3$ minutes that we associate to the WD spin. A confirmation of this association derives from the fact that a few beats with the orbital period are also found in the Lomb-Scargle periodogram (see Fig. \ref{fig_lombscargle} and the discussion in Section 3). This association is also clear in the  $0.1-1$ keV light curve folded at the WD spin period of 20.3 minutes (see Fig. \ref{fig_folded}). Furthermore, with the estimated WD spin, the ratio $P_{spin}/P_{orb}$ is $\simeq 0.09$, a value consistent with that expected for a typical IP system above the period gap. Hence,  we can firmly assign the IP flag to the source VZ Sex.

{VZ Sex also shows a thermal spectrum which is characterized by an intrinsic luminosity of  $\simeq 7\times 10^{31}$ erg s$^{-1}$,} i.e. at the lower end of the distribution of the typical IPs, thus opening to the possibility that a bridge linking the normally bright IPs to the faint population of sources (see recent discussion in  \citealt{mukai2020}) does exist.

\begin{acknowledgements}
 This  paper  is  based on observations from
XMM-Newton, an ESA science mission with instruments
and  contributions  directly  funded  by ESA Member  States  and  NASA. We thank for partial support the INFN projects TAsP and EUCLID. We warmly thank the anonymous Referee for the suggestions that greatly improved the manuscript.

\end{acknowledgements}

{
\software{XSPEC (v12.9.0; \citealt{arnaud})}
\software{SAS (v17.0.0; \citealt{gabriel2004})}
}
%
%

\end{document}